%% file: main.tex
\documentclass[iicol,sn-mathphys,Numbered]{sn-jnl}% Math and Physical Sciences Reference Style
\usepackage{graphicx}
\usepackage{amsmath,amssymb,amsfonts}
\usepackage[svgnames]{xcolor}
\usepackage{manyfoot}
\usepackage{booktabs}
\usepackage{algorithm}
\usepackage{changes}
\usepackage{nicefrac}
\usepackage{afterpage}
\usepackage{orcidlink}

\begin{document}

\input{comments.tex}

\title{Combining lattice QCD and phenomenological inputs on generalised parton distributions at moderate skewness}

\sloppy


\author*[1]{\fnm{Michael Joseph} \sur{Riberdy}\,\orcidlink{0000-0001-6105-5986}}\email{michael.riberdy@cea.fr}
\author[1,2]{\fnm{Herv{\'e}} \sur{Dutrieux}\,\orcidlink{0000-0001-8334-4885}}\email{hldutrieux@wm.edu}
\author[1]{\fnm{C{\'e}dric} \sur{Mezrag}\,\orcidlink{0000-0001-8678-4085}}\email{cedric.mezrag@cea.fr}
\author[3]{\fnm{Pawe{\l}} \sur{Sznajder}\,\orcidlink{0000-0002-2684-803X}}\email{pawel.sznajder@ncbj.gov.pl}

\affil[1]{IRFU, CEA, Universit{\'e} Paris-Saclay, F-91191 Gif-sur-Yvette, France}
\affil[2]{Department of Physics, The College of William \& Mary, Williamsburg VA 23187, US}
\affil[3]{National Centre for Nuclear Research (NCBJ), 02-093 Warsaw, Poland}


\abstract{
We present a systematic study demonstrating the impact of lattice QCD data on the extraction of generalised parton distributions (GPDs).
For this purpose, we use a previously developed modelling of GPDs based on machine learning techniques fulfilling the theoretical requirements of polynomiality, a form of positivity constraint and known reduction limits.
A special care is given to estimate the uncertainty stemming from the ill-posed character of the connection between GPDs and the experimental processes usually considered to constrain them, like deeply virtual Compton scattering (DVCS).
Mock lattice QCD data inputs are included in a Bayesian framework to the prior model which is fitted to reproduce the most experimentally accessible information of a phenomenological model by Goloskov and Kroll.
We highlight the impact of the precision, correlation and kinematic coverage of lattice data on GPD extraction at moderate $\xi$ which has only been brushed in the literature so far, paving the way for a joint extraction of GPDs. 
}




\maketitle


\section{Introduction}
\label{sec:intro}
\input{sec_introduction.tex}

\section{Kinematics and GPD Modelling}
\label{sec:kin_gpd_mod}
\input{sec_kin_gpd_mod.tex}

\section{Representation of experimental data and estimation of uncertainties}
\label{sec:experimental_data_and_unc}
\input{sec_ann_model.tex}

\section{Generating mock lattice QCD data}
\label{sec:data_set_generation}
\input{sec_data_set_generation.tex}

\section{Bayesian Reweighting}
\label{sec:bayesian_reweighting}
\input{sec_bayesian_reweighting.tex}

\section{Results of the reweighting}
\label{sec:gpd_replica_reweighting}
\input{sec_gpd_replica_reweighting.tex}

\section{Conclusion}
\label{sec:conclusion}
\input{sec_conclusion.tex}

%\afterpage{\clearpage}

\bmhead{Acknowledgments}
We thank Savvas Zafeiropoulos, Valerio Bertone, Herv\'e Moutarde, Beno\^it Blossier and Kostas Orginos for stimulating discussions.
This work is supported in part in the framework of the GLUODYNAMICS project funded by the "P2IO LabEx (ANR-10-LABX-0038)" in the framework "Investissements d’Avenir" (ANR-11-IDEX-0003-01) managed by the Agence Nationale de la Recherche (ANR), France.
HD is supported in part by the U.S.~DOE Grant \mbox{\#DE-FG02-04ER41302} and the Gordon and Betty Moore Foundation.
The work of P.S. was supported by the Grant No. 2019/35/D/ST2/00272 of the National Science Centre, Poland

\bibliography{Bibliography}

\end{document}

%% file: comments.tex
\newcounter{comment}
\newcommand{\HDcolor}{blue} 
\newcommand{\CommInlineHD}[1]{{%
		\refstepcounter{comment}%
		\ttfamily\small\textcolor{\HDcolor}{\small[$\blacksquare$ \textbf{\underline{Comment}
$\sharp$\thecomment.}~HD: #1]}}}
\newenvironment{CommBlockHD}%
{\refstepcounter{comment}%
\begin{quote}\renewcommand{\baselinestretch}{1}
\color{\HDcolor}\small$\blacksquare$ \textbf{\underline{Comment} $\sharp$\thecomment.~HD:}}%
{\end{quote}}

\newcommand{\CMcolor}{orange} 
\newcommand{\CommInlineCM}[1]{{%
		\refstepcounter{comment}%
		\ttfamily\small\textcolor{\CMcolor}{\small[$\blacksquare$ \textbf{\underline{Comment}
                    $\sharp$\thecomment.}~CM: #1]}}}
\newenvironment{CommBlockCM}%
{\refstepcounter{comment}%
\begin{quote}\renewcommand{\baselinestretch}{1}
\color{\CMcolor}\small$\blacksquare$ \textbf{\underline{Comment} $\sharp$\thecomment.~CM:}}%
{\end{quote}}

\newcommand{\MRcolor}{ForestGreen} 
\newcommand{\CommInlineMR}[1]{{%
		\refstepcounter{comment}%
		\ttfamily\small\textcolor{\MRcolor}{\small[$\blacksquare$ \textbf{\underline{Comment}
$\sharp$\thecomment.}~MR: #1]}}}
\newenvironment{CommBlockMR}%
{\refstepcounter{comment}%
\begin{quote}\renewcommand{\baselinestretch}{1}
\color{\MRcolor}\small$\blacksquare$ \textbf{\underline{Comment} $\sharp$\thecomment.~MR:}}%
{\end{quote}}

\newcommand{\PScolor}{red} 
\newcommand{\CommInlinePS}[1]{{%
		\refstepcounter{comment}%
		\ttfamily\small\textcolor{\PScolor}{\small[$\blacksquare$ \textbf{\underline{Comment}
$\sharp$\thecomment.}~PS: #1]}}}
\newenvironment{CommBlockPS}%
{\refstepcounter{comment}%
\begin{quote}\renewcommand{\baselinestretch}{1}
\color{\PScolor}\small$\blacksquare$ \textbf{\underline{Comment} $\sharp$\thecomment.~PS:}}%
{\end{quote}}

%%%% Change Package %%%%%%%%%%%%%%%%%%%%

\definechangesauthor[color=\CMcolor]{CM}
\definechangesauthor[color=\MRcolor]{MR}

%% file: sec_introduction.tex
Generalised parton distributions (GPDs) were introduced in the 1990s \cite{Mueller:1998fv,Ji:1996ek,Radyushkin:1997ki} and have been since then a very active topic for theoretical and experimental studies (see \emph{e.g.} the review papers \cite{Diehl:2003ny,Belitsky:2005qn,Kumericki:2016ehc} among others). 
This interest is fuelled both by the interpretation of GPDs as 3D number densities of quarks and gluons within the nucleon \cite{Burkardt:2000za,Diehl:2002he} and by their connection to the energy momentum tensor (EMT) \cite{Ji:1996ek,Polyakov:2002yz}.

A recent study \cite{Bertone:2021yyz} pointed out the large magnitude of uncertainty underlying the connection between GPDs and the exclusive processes usually considered to constrain these objects, such as deeply virtual Compton scattering (DVCS) \cite{Radyushkin:1996nd, Ji:1996nm}, timelike Compton scattering (TCS) \cite{Berger:2001xd} or deeply virtual meson production (DVMP) \cite{Radyushkin:1996ru, Collins:1996fb}.
More precisely, the coefficient functions of these processes are not practically invertible, even when taking into account QCD corrections on a large range of energy scale, and despite the constraints brought by Lorentz covariance.
The results of \cite{Bertone:2021yyz} were confirmed in the moderate $\xi$ range in \cite{Moffat:2023svr}.
In this region, which constitutes the bulk of the most precise experimental data on DVCS, QCD evolution provides little help to alleviate the so-called deconvolution problem.
Recently, efforts were conducted to produce GPD models giving a better account of the uncertainty associated to the ill-defined extraction of GPDs from exclusive processes.
The study \cite{Dutrieux:2021wll} shows that when all the theoretical properties of GPDs (including positivity) are taken into account, the deconvolution uncertainties are mostly present in the so-called Efremov-Radyushkin-Brodsky-Lepage (ERBL) kinematic region. 

The ambiguities of the DVCS, TCS and DVMP processes have led to an on-going considerable theoretical and experimental effort to characterise other exclusive processes with richer kinematic structure.
The best known case is double DVCS (DDVCS) \cite{Guidal:2002kt, Belitsky:2002tf} where both the incoming and outgoing photons are deeply virtual.
This additional kinematic degree of freedom allows, at leading order and in a specific kinematic range, the extraction of the ``deconvoluted'' GPDs.
A new analysis performed in \cite{Deja:2023ahc} suggests that DDVCS is measurable in near-future experiments.
Recently, other processes have been put forward such as di-photon production \cite{Pedrak:2020mfm, Grocholski:2022rqj}, photon meson pair production \cite{Boussarie:2016qop, Duplancic:2018bum}, or the more general single diffractive hard exclusive processes \cite{Qiu:2022bpq, Qiu:2022pla}.

In addition to the experimental inputs, a new source of information regarding GPDs has emerged through lattice QCD simulations.
Indeed, new formalisms developed in \cite{Braun:2007wv,Ji:2013dva,Radyushkin:2017cyf,Ma:2014jla} have offered the possibility to connect non-local spacelike Euclidean matrix elements computed on the lattice to lightlike ones through perturbation theory.
We focus here on two different formalisms. First, the large-momentum effective theory \cite{Ji:2013dva}, known as the ``quasi-distribution'' formalism, has triggered a lot of interest, both in terms of simulation and perturbative computation regarding the so-called matching kernels.
GPD matching kernels have been computed \cite{Ji:2015qla,Liu:2019urm} and the first lattice QCD studies of GPDs have been performed \cite{Alexandrou:2020zbe}.

The second formalism, and the one we will be focusing on through this paper, is known as ``short-distance factorisation'' or ``pseudo-distribution'' formalism \cite{Radyushkin:2017cyf}.
Based on a Lorentz decomposition of the Euclidean matrix elements, it allows the extraction of parton distributions in Ioffe time $\nu$, that is the Fourier conjugate of the momentum fraction $x$.
This method was first introduced in \cite{Radyushkin:2017cyf}, and steady progresses have been done since then, both on the perturbative matching side \cite{Radyushkin:2019owq,Li:2020xml,Ji:2022thb}, and on the lattice simulation one (see for instance \cite{Orginos:2017kos,Egerer:2021ymv}).
The formalism has two advantages: working with ratios of matrix elements greatly simplifies the renormalisation procedure and allows an easier extrapolation to the continuum limit.
It also presents continuous matching kernels.
However, this method - like any lattice-QCD attempt to access lightcone distributions - comes with a difficulty in the actual reconstruction of the momentum dependence of the $\overline{MS}$ distribution.
As only a restricted range of Ioffe times can be probed numerically with acceptable noise levels, the inversion of the Fourier transform to reconstruct an $x$-dependent distribution is an ill-posed inversion problem, also known as an imputation problem.
Several attempts have aleady been performed to try to handle this specific ill-posed problem \cite{Karpie:2019eiq,DelDebbio:2020rgv,Karpie:2021pap}.

The combination of phenomenological and non-local spacelike lattice inputs on parton distributions has already been explored in some recent papers, for instance on the proton PDF \cite{Bringewatt:2020ixn} and pion PDF \cite{JeffersonLabAngularMomentumJAM:2022aix}.
For GPD studies, the situation is made significantly more complicated due to the higher dimensionality compared to ordinary PDFs.
The recent works in \cite{Guo:2022upw, Guo:2023ahv} have associated lattice data with various phenomenological and experimental inputs, where the authors have mostly considered GPD lattice data at vanishing skewness. 
An additional difficulty in offering a framework for the joint study of experimental and lattice inputs on GPDs is the number of parameters involved.
Indeed, for only two light quark flavours, \cite{Guo:2023ahv} needs to model 20 different GPDs -- two valence quark distributions, two sea quark, plus a gluon distribution, repeated for the four helicity combinations $H$, $E$, $\tilde{H}$ and $\tilde{E}$.
This abundance of functions to extract forced the authors to employ only basic modelling of the skewness relevant in the small $\xi$ regime. 

In the present work, we develop a different strategy to combine phenomenology and lattice data focusing on moderate skewness.
In this domain, lattice computations offer the perspective of a significant reduction of the uncertainty associated to the deconvolution problem of the usually considered exclusive processes.
We use a previously developed GPD model presented in \cite{Dutrieux:2021wll} which offers significant flexibility precisely at moderate skewness.
It is illusory in the current state of experimental and lattice data to perform a satisfactory flavour and helicity decomposition with this kind of flexible model.
Therefore, instead of adjusting it on actual experimental and lattice results, we assume that the flavour and helicity decomposition has been obtained already.
On the experimental side, we use the phenomenological information encoded in the Goloskokov and Kroll (GK) model \cite{Goloskokov:2005sd, Goloskokov:2007nt, Goloskokov:2009ia}, whereas on the lattice side, we generate mock lattice data.
Some tension between lattice and experimental data is hinted at in \cite{Bringewatt:2020ixn} and \cite{Guo:2023ahv}, whereas \cite{JeffersonLabAngularMomentumJAM:2022aix}, using short-distance factorization, states that when taking into account all sources of systematic uncertainty, lattice and experimental data are generally compatible.
We therefore use this assumption in the study, which offers the possibility to merge phenomenology and lattice inputs thanks to a Bayesian reweighting procedure.
In short, we probe whether experimental data which suffers from a deconvolution problem and lattice data which suffers from an imputation one can be combined advantageously.
For this first study, we disregard the evolution properties of GPDs and thus the matching kernels between pseudo-Ioffe distributions and GPDs.
The impact of the perturbative QCD corrections is left for future works.

In section \ref{sec:kin_gpd_mod}, we briefly review key aspects of GPDs and their connection with Ioffe-time distributions.
We also revisit the main characteristics of the flexible GPD model introduced in \cite{Dutrieux:2021wll}.
In section \ref{sec:experimental_data_and_unc}, we show how the model is adjusted to a phenomenological parametrization so as to reproduce the main experimental features of a GPD extraction.
Section \ref{sec:data_set_generation} is dedicated to the generation of the set of mock lattice QCD data.
We introduce in section \ref{sec:bayesian_reweighting} the reweighting procedure to combine our mock lattice QCD data to the GPD model.
Finally, we discuss the results of the reweighting in section \ref{sec:gpd_replica_reweighting}, and conclude.

%%% Local Variables:
%%% mode: latex
%%% TeX-master: "main"
%%% End:

%% file: sec_kin_gpd_mod.tex
\subsection{Definition and properties of GPDs}
GPDs can be formally defined as the Fourier transform of non-local matrix elements evaluated on a lightlike distance $z^-$ \cite{Diehl:2003ny}:
\begin{align}
& p^{+}\int\frac{dz^{-}}{2\pi}e^{ixp^{+}z^{-}}\left\langle p_{2}\bigg|\bar{q}\left(-\frac{z}{2}\right)\gamma^{+}q\left(\frac{z}{2}\right)\bigg|p_{1}\right\rangle_{z^{+}
  =|\vec{z}_{\perp}|=0} \nonumber \\
  &= H^{q}(x,\xi,t)\bar{u}(p_{2})\gamma^{+}u(p_{1}) \nonumber \\
  &\hspace{20pt} \quad +E^{q}(x,\xi,t)\bar{u}(p_{2})\frac{\sigma^{+\alpha}\Delta_{\alpha}}{m}u(p_{1}),
\end{align}
where we have restricted here our definition to the unpolarized quark GPD in a nucleon whose mass is labelled $m$.
We also omitted the Wilson line in the definition of the operator, that needs to be added when working in gauges other than lightcone one.
The average momentum of the incoming and outgoing hadronic states $|p_{1}\rangle$ and $|p_{2}\rangle$ is defined as:
\begin{equation}
p=\frac{p_{1}+p_{2}}{2},
\end{equation}
while $x$ is the usual lightcone momentum fraction defined as:
\begin{equation}
x=\frac{k_{1}^{+} + k_{2}^{+}}{2p^{+}}
\end{equation}
with $k^{+}_{1,2}$ the incoming and outgoing momentum of the struck quark.
The total four-momentum transfer squared $t$ is given by:
\begin{equation}
t=(p_{1}-p_{2})^{2}
\end{equation}
and the skewness $\xi$ is defined as:
\begin{equation}
\xi=\frac{p_{1}^{+}-p_{2}^{+}}{2p^{+}}.
\end{equation}
GPDs have to obey several properties which play a crucial role in their modelling.
First, the forward limit of GPDs is given in terms of PDFs, such that:
\begin{equation}
  \label{eq:ForwardLimit}
  H^q(x,0,0) = q(x)\Theta(x) - \bar{q}(-x)\Theta(-x)
\end{equation}
where $q(x)$ is the unpolarised quark PDF of flavour $q$ and $\Theta$ the Heaviside step function.
GPDs are also connected to electromagnetic form factors (EFFs) through:
\begin{equation}
  \label{eq:EFFSumRules}
  \int_{-1}^1\textrm{d}x H^q(x,\xi,t) = F^q_1(t),
\end{equation}
where $F^q_1$ is the contribution of the quark flavour $q$ to the Dirac EFF.
GPDs have to obey the so-called polynomiality property, stating that their Mellin moments are polynomials in $\xi$ of a given order \cite{Ji:1998pc,Radyushkin:1998bz}:
\begin{align}
  \int_{-1}^1 \textrm{d}x\, x^n H^q(x,\xi,t) &= \sum_{i=0}^{\left[\frac{n}{2}\right]}\xi^{2i}A^q_{n,2i}(t) \nonumber\\
  &+ \textrm{mod}(n,2)C_n(t)\xi^{n+1},
  \label{eq:polynom}
\end{align}
where $[\dots]$ is the floor function and $\textrm{mod}(n,2)$ is 0 for $n$ even, and 1 otherwise.
The polynomiality property is understood as a consequence of the Lorentz covariance of GPDs, and can be systematically implemented thanks to the Radon transform \cite{Teryaev:2001qm,Chouika:2017dhe,Mezrag:2022pqk} in the double distribution formalism \cite{Radyushkin:1998bz}.
Finally, let us highlight that GPDs are also constrained by the so-called positivity properties \cite{Radyushkin:1998es,Pire:1998nw,Diehl:2000xz,Pobylitsa:2002gw}, the most classical one being given by:
\begin{align}
  \label{eq:Positivity}
  &\left|H^q(x,\xi,t) \right| \le \sqrt{\frac{q\left(\frac{x+\xi}{1+\xi}\right)q\left(\frac{x-\xi}{1-\xi}\right)}{1-\xi^2}}.
\end{align}
Respecting all these constraints at once represents a challenge for the phenomenology of GPDs.

GPDs can also be expressed as a function of the Ioffe time parameter $\nu = p\cdot z$ \cite{Braun:1994jq}, which is the Fourier conjugate of the average momentum fraction $x$.
The relation between $\hat{H}(\nu,\xi,t)$ and $H(x,\xi,t)$ is thus given by \cite{Radyushkin:2019owq}:
\begin{align}
  \hat{H}^q(\nu,\xi,t) = \int_{-1}^1 \textrm{d}x e^{ix\nu} H^q(x,\xi,t)
  \label{eq:defioffetime}
\end{align}
In the following, we drop the ``hat'' on $\hat{H}(\nu)$ as the explicit dependence in $\nu$ or $x$ is enough to indicate whether we work in Ioffe time or momentum space. 

As we have mentioned before, non-local matrix elements computed on the lattice are Euclidean, so the distance $z$ between the operators defining the GPD is spacelike ($z^2 < 0$) contrary to the lightlike distance $z^-$ used in the usual lightcone definition of parton distributions.
As a result, Ioffe-time pseudo-distributions computed on the lattice, once the continuum limit has been appropriately taken, are functions $\mathfrak{M}(\nu, \xi, t, z^2)$, which can be matched perturbatively to the $\overline{MS}$ scheme to yield the Ioffe-time distributions such as $H(\nu, \xi, t, \mu^2)$ (see \cite{Radyushkin:2019owq, Ji:2022thb}).
Matching from $z^2$-dependent pseudo-distributions to $\mu^2$-dependent $\overline{MS}$ distributions is only a minor correction at small Ioffe time, which we will neglect in this paper.
This effect should however be properly accounted for in further studies. 

In the following, for simplicity we limit the scope of our study to the singlet sea quark GPD $H^{q(+)}(x, \xi, t) = H^{q}(x, \xi, t) -  H^{q}(-x, \xi, t)$.
Imposing this parity property means that we will only study the imaginary part of $H(\nu)$.
We also ignore the $t$-dependence of GPDs and the entanglement between $t$ and $\xi$ which splits the kinematic domain between physical and unphysical regions, to work with $t=0$ all along.
This choice is made to primarily focus on the deconvolution of the $x$ and $\xi$ dependence helped by lattice data.
The implications of the $t$ dependence are left for a future work.  

\subsection{GPD Modelling with artificial neural networks}
\label{sec:ANNMod}

Our study is based on methods recently developed in \cite{Dutrieux:2021wll}, where the use of artificial neural network (ANN) techniques in a direct modelling of GPDs was proposed for the first time.
This new way of modelling has been designed to address the problem of model dependence and implementation of the theoretical constraints one encounters in the GPD phenomenology, but also to facilitate a future inclusion of lattice QCD information.
To keep our article self-consistent and self-contained, we remind now important details on ANN GPD modelling. 

Our GPD model based on ANNs significantly differs from a textbook implementation of machine learning techniques (see e.g. Ref.~\cite{Cybenko}).
The reason for that is the desire to fulfil in the architecture of the neural network a set of theory-driven constraints for a valid GPD model.
These constraints are among others linked to the parity properties of GPDs, the polynomiality property \eqref{eq:polynom} and known limits like \eqref{eq:ForwardLimit}.
The positivity constraints \eqref{eq:Positivity} are not guarantied at the level of the architecture of the network, but rather enforced numerically during the training procedure.  

The model proposed in Ref.~\cite{Dutrieux:2021wll} is built in the double distribution space involving variables $\beta$ and $\alpha$.
The relation with the GPD $H$ in momentum space is given by the Radon transform:
\begin{equation}
  \label{eq:DDFormalism}
	H(x,\xi,t) = \int \mathrm{d}\Omega\,  F(\beta, \alpha,t) \,,
\end{equation}
where $\mathrm{d}\Omega =\mathrm{d}\beta\,\mathrm{d}\alpha\,\delta(x-\beta-\alpha\xi)$ and $|\alpha|+|\beta|\leq 1$. In \eqref{eq:DDFormalism}, $F(\beta, \alpha,t)$ is called a double distribution and is the object we will model. The benefit of using the double distribution space is an automatic fulfilment of the polynomiality property by the resulting GPD $H(x, \xi,t)$.

To achieve a satisfactory flexibility and reproduction of known limits, our double distribution model is composed of three parts:
\begin{equation}
  \label{eq:DDComplete}
	(1-x^2)  F_{C}(\beta, \alpha) + 
    	(x^2 - \xi^2) F_{S}(\beta, \alpha) + 
    	\xi F_{D}(\beta, \alpha)\,.
\end{equation}
 Let us address successively the role of each term in \eqref{eq:DDComplete}. The first one, 
\begin{equation}
\label{eq:fC}
	F_{C}(\beta, \alpha) = f(\beta) h_{C}(\beta, \alpha)\frac{1}{1-\beta^{2}},
\end{equation}
ensures by design the proper reduction to the forward limit and has the necessary flexibility to model the $x=\xi$ line, which is particularly relevent for the current GPD phenomenology.
The prefactor $(1-x^2)$ of $F_C(\beta, \alpha)$ in \eqref{eq:DDComplete} combined with $1/(1-\beta^2)$ in \eqref{eq:fC} was introduced to facilitate the fulfilment of the positivity constraint \eqref{eq:Positivity}.
The forward limit (unpolarised PDF for the GPD $H$) is denoted by $f(\beta) \equiv H(\beta,0,0)$, while $h_{C}(\beta, \alpha)$ is a profile function generalising that proposed by Radyushkin \cite{Radyushkin:1998es}.
In this study it is given by:
\begin{equation}
	h_{C}(\beta, \alpha) = \frac{\mathrm{ANN}_{C}(|\beta|, \alpha)}{\displaystyle\int_{-1+|\beta|}^{1-|\beta|} \mathrm{d}\alpha \mathrm{ANN}_{C}(|\beta|, \alpha) }  \,.
\end{equation}
Thanks to a special design of the activation function and the use of the absolute value, the neural network $\mathrm{ANN}_{C}(|\beta|, \alpha)$ is even w.r.t. both $\beta$ and $\alpha$ variables, and it vanishes at the edge of the support region $|\beta| + |\alpha| = 1$.
The symmetry in $\beta$ is introduced to keep the resulting GPD an odd function of $x$ (relevant for phenomenology of DVCS and TCS), while the symmetry in $\alpha$ is mandatory, as it allows fulfillment of the time reversal property, i.e. the invariance under $\xi \leftrightarrow -\xi$ exchange.
The normalisation by the integral over $\alpha$, which can be done analytically, allows enforcement of the proper forward limit, while the rest of the model is typically trained to reproduce the diagonal $x = \xi$ probed by amplitudes of processes like DVCS, TCS and DVMP. 

As the term $F_{C}(\beta, \alpha)$ was found to be tightly constrained by the necessity of reproducing both $\xi=0$ and $x=\xi$ lines, an additional term $F_{S}(\beta, \alpha)$ was introduced.
This term explicitly vanishes on the $\xi=0$ and $x=\xi$ lines, i.e. it does not contribute to the fit of $F_{C}(\beta, \alpha)$ on the phenomenological inputs.
Rather, $(x^2-\xi^2)F_S(\beta, \alpha)$ represents a contribution that is entirely unconstrained by LO DVCS data and the knowledge of PDFs, and aims at reproducing the deconvolution uncertainty of exclusive processes.
Precisely, it corresponds to a LO shadow distribution as defined and studied in \cite{Bertone:2021yyz}.
$F_S$ is constructed in the following way:
\begin{equation}
  \label{eq:ModelShadow}
	F_{S}(\beta, \alpha) = f(\beta) h_{S}(\beta, \alpha) \,.
\end{equation}
where:
\begin{flalign}
  \label{eq:ModelShadowhS}
	h_{S}(\beta, \alpha) / N_{S} & = \frac{\mathrm{ANN}_{S}(|\beta|, \alpha)}{\displaystyle\int_{-1+|\beta|}^{1-|\beta|} \mathrm{d}\alpha \mathrm{ANN}_{S}(|\beta|, \alpha) } \nonumber \\
& - \frac{\mathrm{ANN}_{S'}(|\beta|, \alpha)}{\displaystyle\int_{-1+|\beta|}^{1-|\beta|} \mathrm{d}\alpha \mathrm{ANN}_{S'}(|\beta|, \alpha) }
\,.
\end{flalign}
Since this contribution is not constrained in the fit, the major limit on its size, aside from the maximal flexibility of the neural network, is really imposed by the positivity constraint.
During training, we seek to maximise the $N_{S}$ normalisation factor in \eqref{eq:ModelShadowhS} within the limits allowed by positivity so as to leverage the maximal flexibility.
Writing the function $h_S(\beta, \alpha)$ as the difference of two different profile functions characterized by $\mathrm{ANN}_{S}(|\beta|, \alpha)$ and $\mathrm{ANN}_{S'}(|\beta|, \alpha)$ ensures that $F_S(\beta, \alpha)$ brings no contribution to the forward limit.
The $f(\beta)$ factor helps to enforce the positivity.

Finally, $F_{D}(\beta, \alpha)$ gives the additional flexibility necessary to model the $D$-term, a degree of freedom of GPDs associated to the final terms in $\xi^{n+1}$ in \eqref{eq:polynom} and which plays a crucial role in the characterisation of partonic matter \cite{Goeke:2007fp, Polyakov:2018zvc}.
One has :
\begin{equation}
F_{D}(\beta, \alpha) =  \delta(\beta) D(\alpha) \,,
\end{equation}
and
\begin{equation}
D(\alpha) = (1-\alpha^2) \sum_{\substack{i=1 \\ \mathrm{odd}}}^{N} d_{i}C_{i}^{\nicefrac{3}{2}}\left(\alpha\right) \,,
\end{equation}
where $d_{i}$ are coefficients of the expansion of $D$-term into Gegenbauer polynomials, and where $N$ is an arbitrary truncation parameter.

%%% Local Variables:
%%% mode: latex
%%% TeX-master: "main"
%%% End:

%% file: sec_ann_model.tex
Let us now discuss how the model encodes a representation of experimental data for processes like DVCS, TCS and DVMP which we will use as an input for the lattice QCD impact study. We stress again that experimental data for the aforementioned processes mostly probe GPDs at the $x=\xi$ line (with some additional information on the $D$-term), and that at $t = 0$, the forward limit, \textit{i.e.} the PDF, is very well known from a wealth of inclusive and semi-inclusive processes. 

We use a built-in PDF parameterisation proposed in \cite{Goloskokov:2006hr} for $f(\beta)$ involved in \eqref{eq:fC} and \eqref{eq:ModelShadow}. Free parameters of the $F_{C}(\beta, \alpha)$ term are constrained by $200$ points of $H^q(x,\xi=x)$ generated with the GK model \cite{Goloskokov:2005sd, Goloskokov:2007nt, Goloskokov:2009ia} spanning over the range of $-4 < \log_{10}(x = \xi) < \log_{10}(0.95)$ at fixed $t = 0$ and $\mu^2 = 4~\mathrm{GeV}^{2}$. The term $F_{S}(\beta, \alpha)$ is only constrained by the positivity requirement \eqref{eq:Positivity}, giving rise to large uncertainties when $x < \xi$. The details of the constraining procedure are given in Ref.~\cite{Dutrieux:2021wll}.

The uncertainty of the model is encoded in a collection of $101$ replicas. This way is convenient for propagation of uncertainty to any related quantity and for the use of Bayesian reweighting techniques. A single replica represents the outcome of the independent fit to $200$ $x=\xi$ points indicated in the previous paragraph, with a random choice of the initial parameters (weights and biases of ANNs, and normalisation parameter $N_S$ in \eqref{eq:ModelShadowhS}). To evaluate the mean and standard deviation of a quantity derived from the GPD at a specific kinematic point -- which can be the GPD itself, a 3D number density, an observable, etc. -- one may use any statistical estimator of the mean and uncertainty of the population $X$ of $101$ values returned by replicas. In this analysis we use respectively the median and the median absolute deviation (MAD): 
\begin{equation}
\bar{\sigma}\equiv 1.4826\times\text{median}\bigg(|\text{median}(X)-X|\bigg),
\label{eq:mad}
\end{equation}
where the factor $1.4826$ is derived from the assumption of gaussianity. 

We note that the population of $101$ values returned by replicas may contain outliers, i.e. pathological values with respect to the other entries, distorting the evaluation of mean and uncertainty. This may happen due to instabilities in the numeric procedures, which cannot be entirely avoided due to the complexity of GPD modelling and the constraining procedure. Many methods for detecting and removing outliers were proposed in the literature, like the $3\sigma$-method \cite{10.1145/3310205}. Alternatively, one can chose robust estimators with respect to outliers, our motivation for using MAD.

\begin{figure}[!ht]
\centering
\includegraphics[width=0.8\linewidth]{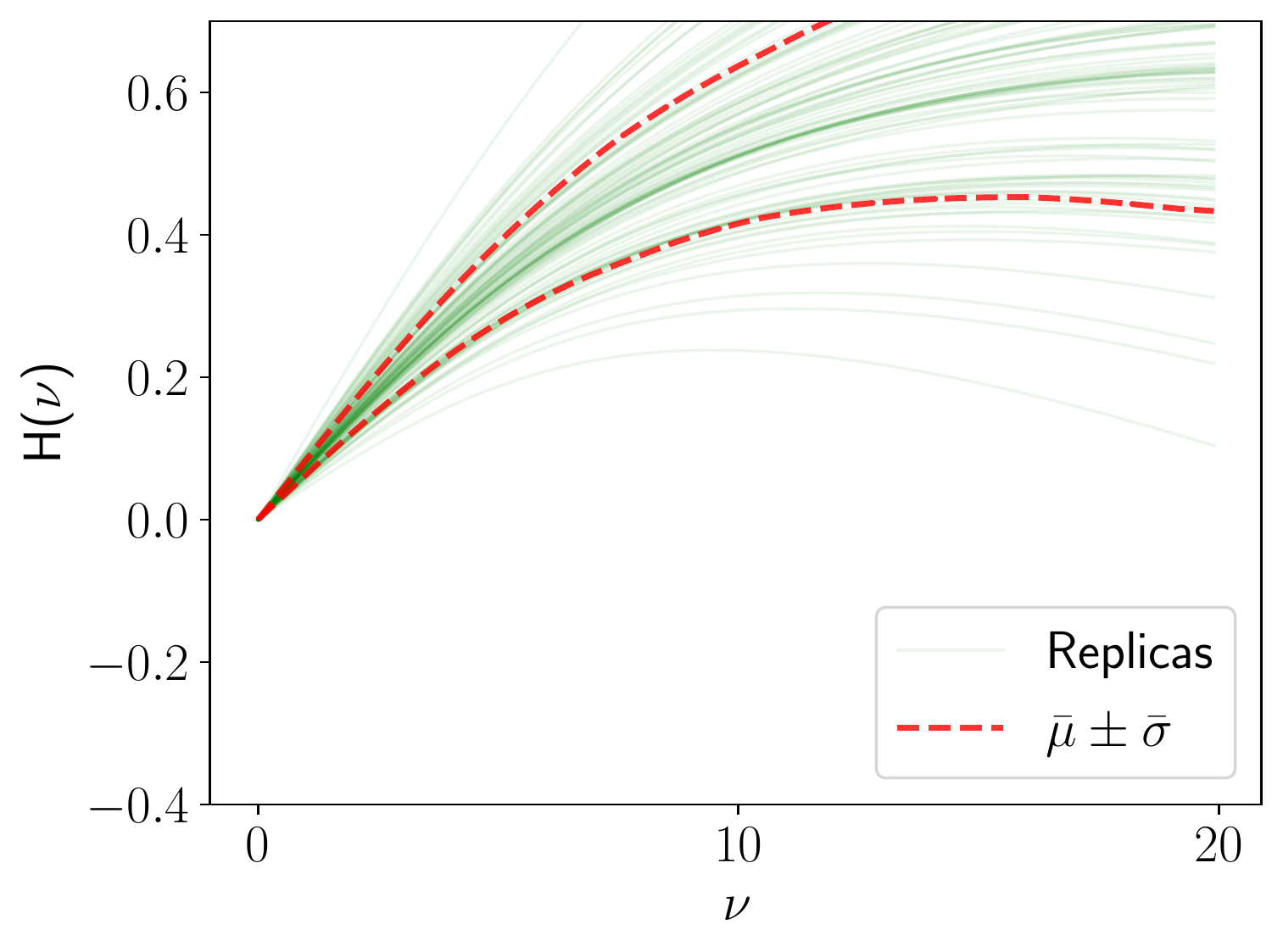}
\includegraphics[width=0.8\linewidth]{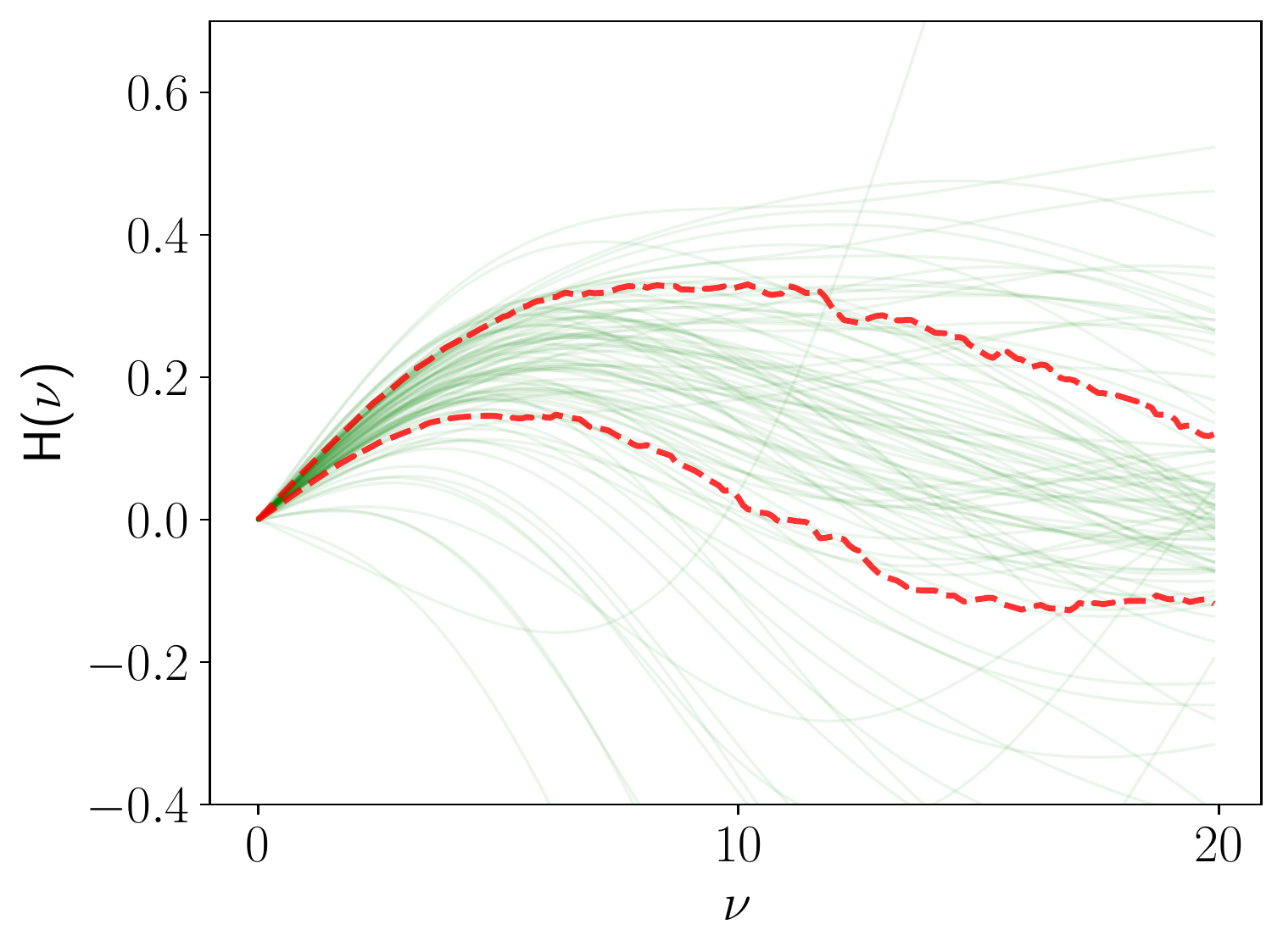}
\includegraphics[width=0.8\linewidth]{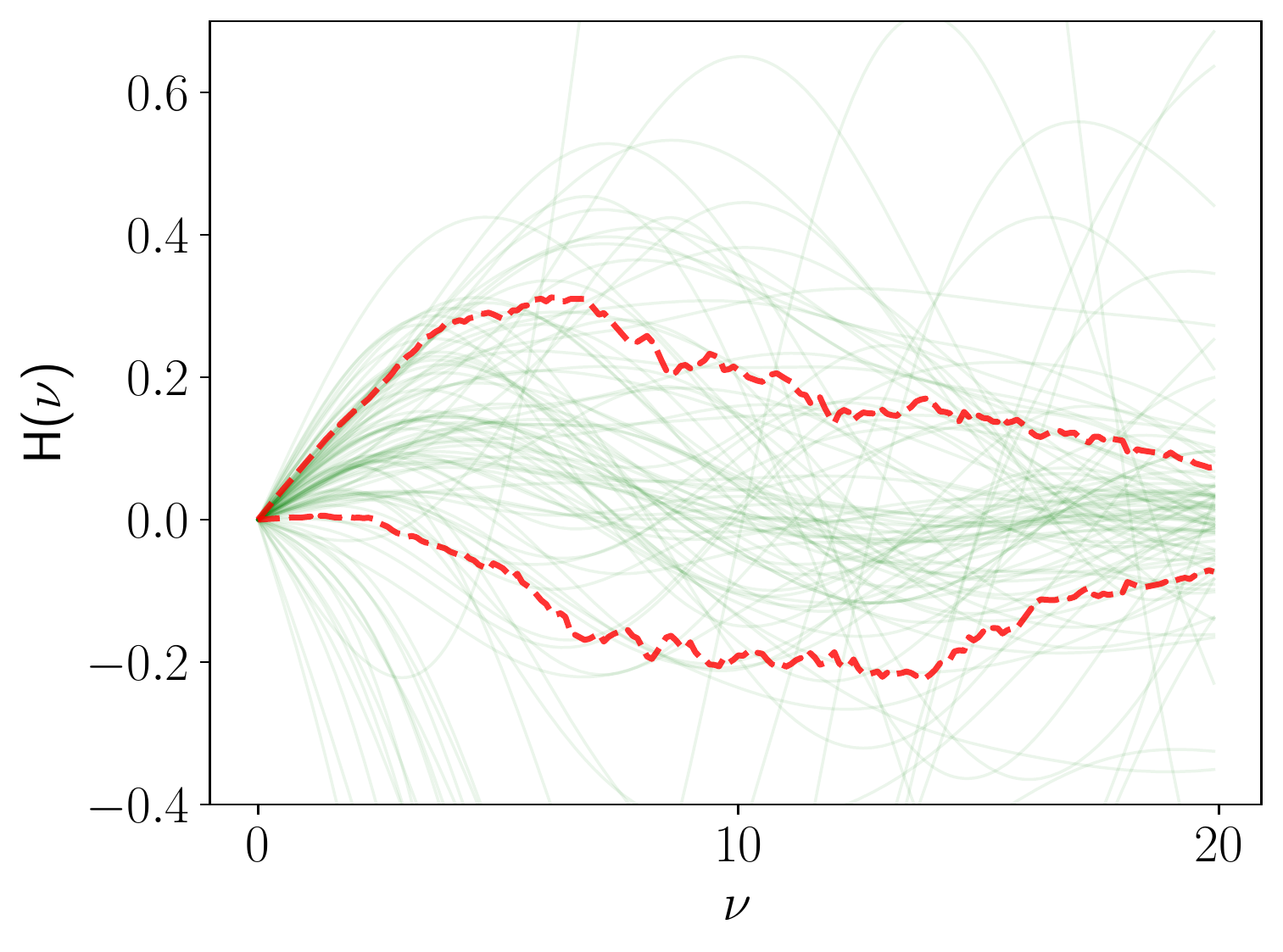}
\caption{The set of GPD replicas $\textrm{Im}\,H(\nu, \xi ,t=0)$ between Ioffe times $\nu=0$ and $\nu=20$ at $\xi=0.1$ (top), $\xi=0.5$ (middle), $\xi=0.9$ (bottom) and their corresponding one standard deviation bands.
Due to waning support $x > \xi$ as $\xi$ increases, the replicas become less constrained by positivity, oscillate more heavily, and decohere.}
\label{fig:bare_reps}
\end{figure}

Figure~\ref{fig:bare_reps} shows $101$ replicas of $H(\nu, \xi, t =0)$ evaluated at three values of $\xi$, together with its $1 \bar{\sigma}$ band. For small values of $\xi$, the replica bundle is extremely coherent, or auto-correlated, at small Ioffe time. This is due to the fact that the positivity constraint limits considerably the freedom of the model for $x > \xi$, on a region that is therefore all the more extended that $\xi$ is small.

%% file: sec_data_set_generation.tex
After describing in the previous section the fit of our flexible GPD model with phenomenological inputs, we turn to the question of the generation of plausible mock lattice QCD data. In accordance with our choice of working in the pseudo-distribution formalism, we will generate data at small Ioffe time $\nu$. More precisely, for each value of $\xi$ that we will consider, we generate the data in three independent batches, corresponding to the following regions in $\nu$:
\begin{enumerate}
    \item $0.2\leq\nu\leq2$, with data spaced regularly with interval $\Delta\nu=0.2$,
    \item $2.2\leq\nu\leq4$, with interval $\Delta\nu=0.2$,
    \item $4.4\leq\nu\leq6$, with interval $\Delta\nu=0.4$.
\end{enumerate}
We introduce these three independent batches of lattice QCD data to loosely mimic actual simulations where various sets of points with different correlations arise from varying the boosted hadron momenta and the separation between fields in the non-local operator. 

We assume the correlation between different batches of data to be zero. On the contrary, we consider the existence of correlations inside the batches, characterized for commodity by a simple coefficient $0\leq c<1$, which is the same between any two lattice data points in the same batch, and does not vary from batch to batch.
This choice of constant correlation coefficient is obviously an oversimplification, but it already allows us to get a rough estimate of the impact of correlations on the determination of GPDs.

\begin{figure}[!ht]
\centering
\includegraphics[width=0.8\linewidth]{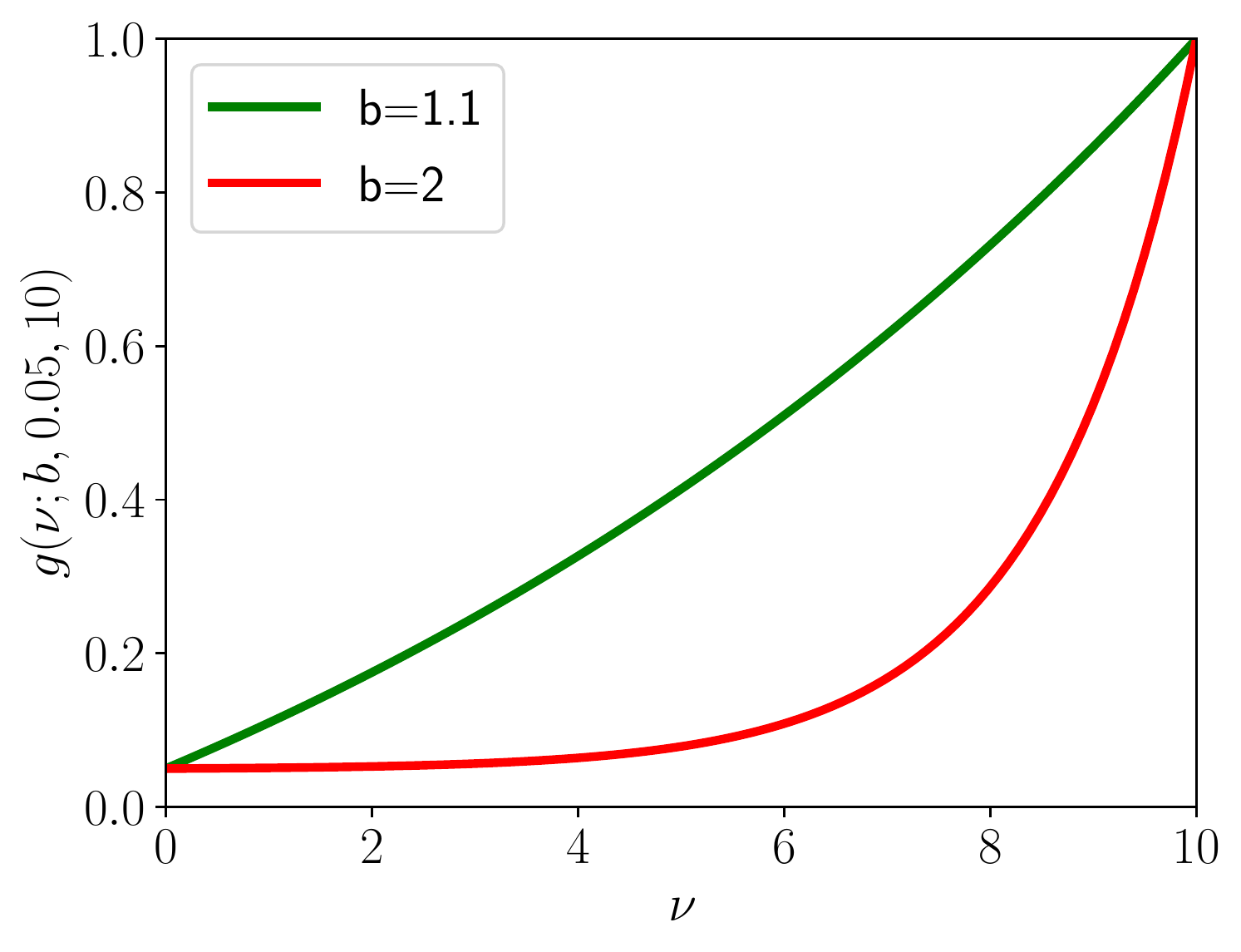}
\caption{The relative uncertainty of the mock lattice data $g(\nu;b,0.05,10)$ is displayed as a function of the Ioffe time $\nu$ for $b=1.1$ (green) and for $b=2$ (red).}
\label{fig:fs}
\end{figure}

We characterise how uncertainty grows from $\nu = 0$ to $\nu = 10$ by an exponential slope parameter called $b$. Figure \ref{fig:fs} presents the relative uncertainty profiles which we will use in this study, given by
\begin{align}
    &g(\nu; b, s = 5\%, \nu_{\textrm{max}} = 10) = \frac{s (b^\nu - b^{\nu_{\textrm{max}}})+1-b^\nu}{1-b^{\nu_{\textrm{max}}}}\,.
\end{align}
As can be immediately checked, this formula guarantees that when $\nu = \nu_{\textrm{max}}$, the relative uncertainty is 100\%, and when $\nu = 0$, it saturates to $s$, here chosen as 5\%.
We choose the uncertainty to reach 100\% at $\nu_{\textrm{max}}=10$ to replicate the common behaviour of lattice data, which tend to exhibit uncertainty on the order of its central value around this point in Ioffe time (see, for example, Fig. 9.a. of \cite{Egerer:2021ymv}). 
The parameter $b$ controls the steepness of the uncertainty increase. For $b \rightarrow 1$, uncertainty increases linearly with Ioffe time, while for larger values of $b$, it starts with a plateau of good precision and then degrades very quickly.
The two cases $b = 1.1$ and $b= 2$ presented on figure \ref{fig:fs} represent two archetypes of lattice uncertainties, corresponding respectively to data of bad or good signal to noise ratio.

We will therefore use as uncertainty of our mock lattice data
\begin{equation}
\sigma^{\textrm{Latt}}(\nu) \equiv g(\nu;b, s = 5\%, \nu_{\textrm{max}} = 10)\bar{\mu}(\nu)\,,
\end{equation}
where $\bar{\mu}(\nu)$ is the median of the output of our flexible GPD model at $\nu$ and the relevant value of $\xi$.
Each of the three batches of data points contains a number of samples at Ioffe times $\nu_{i}$.
We characterize the correlation between these points inside each batch by a coefficient $c$ and express the associated covariance matrix
\begin{equation}
\Omega_{i,i'}^{\textrm{Latt}}\equiv\Big{(}\delta_{i,i'}+(1-\delta_{i,i'})c\Big{)}\sigma^{\textrm{Latt}}(\nu_{i})\sigma^{\textrm{Latt}}(\nu_{i}'),
\end{equation}

We then draw the central values $\mu_i^{\textrm{Latt}}$ of our mock lattice data points in a normal distribution centered on the median of the output of the flexible GPD model at each $\nu_i$, with the covariance matrix $\Omega_{i,i'}^{\textrm{Latt}}$. This means that we choose our mock lattice data to be globally compatible with the central value of our flexible GPD model fitted on phenomenological inputs.

\begin{figure*}[!ht]
\centering
\includegraphics[width=0.8\textwidth]{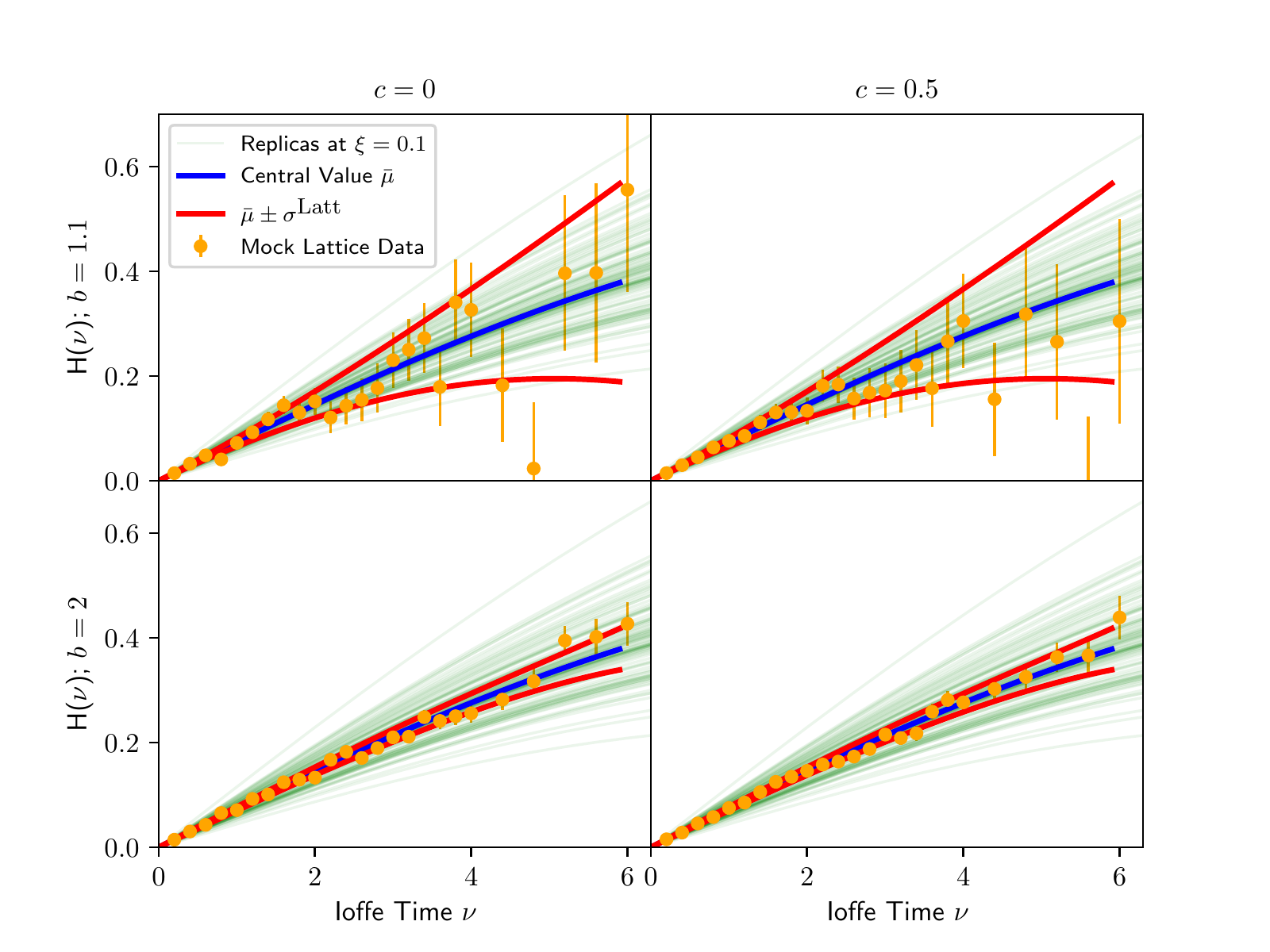}
\caption{The set of GPD replicas between Ioffe times $\nu=0$ and $\nu=6$ at $\xi=0.1$ (green), their median (blue), the $1\sigma$ band (red) corresponding to $b=1.1$ (top) and $b=2$ (bottom), and the corresponding generated mock lattice data set (orange) determined by $c=0$ (left) and $c=0.5$ (right) are shown.}
\label{fig:l_gen}
\end{figure*}

Figure \ref{fig:l_gen} gives an example of mock lattice data set (orange points) superposed to the replicas of our GPD model. The four panels demonstrate the effect of various combinations of correlation coefficient $c$ and level of noise parameter $b$. Increasing $c$ increases the degree to which one of the central values influences the choice of the others within a given batch in $\nu$, while increasing the parameter $b$ results in a data set much more concentrated around the maximum likelihood of the GPD model.

%%% Local Variables:
%%% mode: latex
%%% TeX-master: "main"
%%% End:

%% file: sec_bayesian_reweighting.tex
Now, we would like to assess the impact of the different sets of mock lattice data generated in the previous section on the GPD model. For this purpose, for each replica $R_k$, we introduce a goodness-of-fit estimator $\chi^{2}_{k}$:
\begin{eqnarray}
\chi^{2}_{k} &=& \sum_{\textrm{batches}}\sum_{i,i'}(\mu_{i}^{\textrm{Latt}}-R_{k}(\nu_{i}))\Big{(}(\Omega^{\textrm{Latt}})^{-1}\Big{)}_{i,i'}
\nonumber\\
&\times& (\mu_{i'}^{\textrm{Latt}}-R_{k}(\nu_{i'})),
\end{eqnarray}
where $\mu_{i}^{\textrm{Latt}}$ is the central value of the lattice data generated at point $\nu_{i}$. With this definition, $\chi^{2}_{k}$ estimates the relative compatibility of a given replica $R_{k}$ with the mock data within the uncertainty of the latter. Bayesian reweighting consists in affecting to each replica $R_k$ a weight $\omega_k$ expressed from the goodness-of-fit $\chi^2_k$ through \cite{Ball:2011gg}
\begin{equation}
\omega_{k} =  \frac{(\chi^{2}_{k})^{\frac{N-1}{2}}}{Z}e^{-\frac{\chi^{2}_{k}}{2}},
\end{equation}
where $Z$ is a normalization factor such that $\sum_{k}\omega_{k}=1$. 
One can further define the effective fraction of replicas $R$ which are compatible with the new data set as
\begin{equation}
\tau\equiv\frac{\exp(\sum_{k}\omega_{k}\ln(\omega_{k}))}{K}, \label{eq:tau}
\end{equation}
where the exponentiated value is the Shannon entropy of the weight set.

In the following, we will generate several sets of mock lattice data at various values of $\xi$ and reweight our GPD model altogether with these new data at several $\xi$. We call this the ``multikinematic'' reweighting, as opposed to the ``monokinematic'' reweighting where only one value of $\xi$ is considered. As we do not model any correlation between mock lattice data at different values of $\xi$, the total $\chi^2$ will then be the sum of the $\chi^2$ evaluated at each $\xi$.

\subsection{Reweighted statistics}

Once weights $\omega_k$ have been attributed to the replicas, we need to define the central value and standard deviation of a weighted distribution. At each kinematic where we want to evaluate the weighted central value and deviation, we first order the replicas monotoneously, and then define the weighted median $\bar{\mu}_\omega$ as the element satisfying the condition
\begin{eqnarray}
\sum_{k=1}^{l-1}\omega_{k}\leq\frac{1}{2} \quad \textrm{and} \quad \sum_{k=l+1}^{k_{\textrm{max}}}\omega_{k'}\leq\frac{1}{2}.
\end{eqnarray}
The weighted median is the element which most accurately splits the weights evenly. The weighted standard deviation estimator $\bar{\sigma}_\omega$ is obtained by simply replacing the median by its weighted equivalent in \eqref{eq:mad}.

\subsection{Metrics of the impact of Bayesian reweighting}

The effective fraction of replicas compatible with the data set $\tau$ defined in \eqref{eq:tau} tells us how constraining the new data set is compared to the prior model. It is mostly a tool to evaluate the statistical significance of the reweighted distribution: if $\tau$ is too small, the weighted mean and standard deviation should not be considered as statistically significant. However, $\tau$ brings only indirect information on the reduction of uncertainty. To characterise the latter, which is our main physical objective, we introduce the ratio $\Sigma(y)$, such as 
\begin{equation}
\Sigma(y) \equiv \frac{\bar{\sigma}_\omega(y)}{\bar{\sigma}(y)}.
\end{equation}
At a given value of $y$ (which can represent any kinematic, in momentum space or Ioffe time), it represents the fraction of uncertainty remaining after reweighting. If $\Sigma(y)=1$, the uncertainty of the replica bundle at $y$ has not decreased via reweighting. If $\Sigma(y)<1$, some reduction of uncertainty has occurred via reweighting at that point. We further define the average retainment of uncertainty in Ioffe time as
\begin{equation}
r_{\nu}=\frac{1}{\nu_{\max} - \nu_{\min}}\int_{\nu_{\min}}^{\nu_{\max}} d\nu\,\Sigma(\nu),
\end{equation}
and a corresponding ratio in momentum space where we adopt a logarithmic scale,
\begin{equation}
r_{\textrm{ln}x}=\frac{1}{\textrm{ln}(x_{\max}/x_{\min})}\int_{x_{\min}}^{x_{\max}} \frac{dx}{x}\Sigma(x).
\end{equation}
We calculate the uncertainty retainment values $r_{\nu}$ for the region $0\leq\nu\leq20$ and $r_{\textrm{ln}x}$ for the region $5\times 10^{-3}\leq x\leq1$. These two metrics, intended for $\nu$ and $x$ spaces respectively, assign a global numerical value to the reduction of uncertainty following the introduction of the mock lattice data, which will be convenient to compare various scenarios.

One should note that although we generate mock lattice data in the range $0\leq\nu\leq 6$, our uncertainty retainment $r_{\nu}$ metric extends up to $\nu = 20$.
We do this to assess the ability for lattice data to decrease the uncertainty of our GPD model even in Ioffe time regions where we do not expect to be able to extract statistically significant lattice data. Indeed, \cite{JeffersonLabAngularMomentumJAM:2022aix} concluded that the data at low values of Ioffe time, thanks to their smaller uncertainties, represented in effect the bulk of the constraint even at larger Ioffe times. 
We also do not include the entire region $0\leq x\leq 1$ in our metric of uncertainty retainment $r_{\textrm{ln}x}$, integrating only from the lower bound $x=5\times10^{-3}$.
We choose to cap the integration with this lower bound given that $\Sigma(x)$ becomes relatively constant at lower $x$.
The inclusion of such behaviour in the integration region would completely dominate the discrimination effects at large $x$.

%%% Local Variables:
%%% mode: latex
%%% TeX-master: "main"
%%% End:

%% file: sec_gpd_replica_reweighting.tex
\subsection{Monokinematic Reweighting}
Let us now apply the tools of Bayesian reweighting using our GPD model fitted on phenomenological inputs as a prior, and our mock data as the new information.
As a first example, we look at monokinematic reweighting, \textit{i.e.} we add mock data at a single value of $\xi$ and measure its impact on the GPD extraction at this same value. We recall that, on average, the generated mock lattice data becomes closer to the most likely output of the prior model as $b$ increases. As $c$ increases, the mock lattice data remains more consistently above or below the central value of the prior model. 

\begin{figure*}
\centering
\includegraphics[width=1\textwidth]{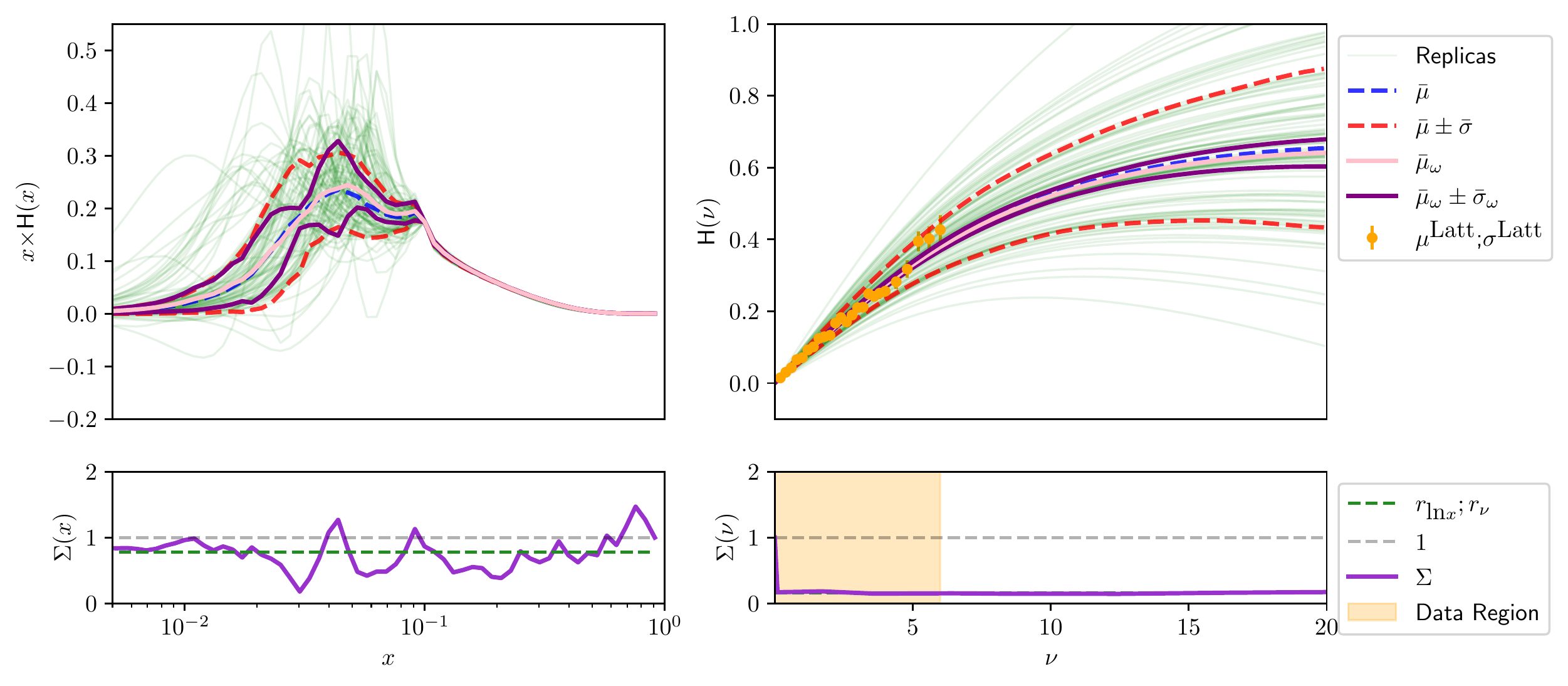}
\caption{Upper plots: The set of GPD replicas at $\xi=0.1$ in momentum space (green, left) and Ioffe times space (green, right), their central value (pink-solid), one $\sigma$ band (purple-solid), and the reweighted central value (blue-dashed) and reweighted one $\sigma$ band (red-dashed).
The mock lattice data was generated using $b=2$ (high precision) and $c=0$ (no correlation) at $\xi=0.1$ (orange-dotted, right).
Lower plots: The reweighted to initial uncertainty ratio (purple-solid), the average uncertainty retainment in both $x$ ($r_{\textrm{ln}x}=0.78$) and $\nu$ ($r_{\nu}=0.16$) (green-dashed), and the range in which the lattice data was generated $\nu=0$ to $\nu=6$ (orange-shade, right).
The associated effective fraction of replicas retained after reweighting is given by $\tau=0.3$.}
\label{fig:in_0.1_out_0.1_cor_0_b_2_rewe}
\end{figure*}

\begin{figure*}
\centering
\includegraphics[width=1\textwidth]{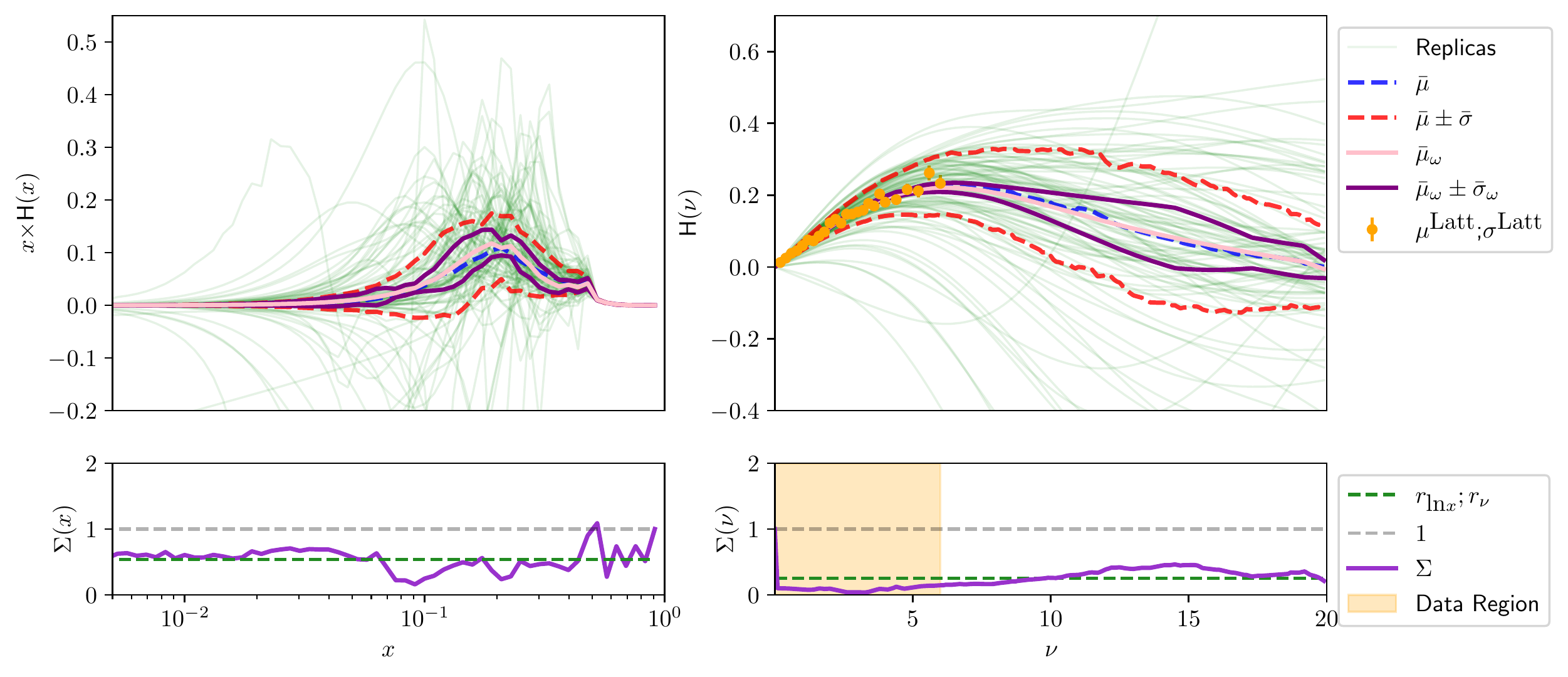}
\caption{Same caption as figure \ref{fig:in_0.1_out_0.1_cor_0_b_2_rewe} up to the fact that the GPD is shown at $\xi = 0.5$, with mock data added at $\xi = 0.5$ and similarly $b=2$ (high precision) and $c=0$ (no correlation). The average uncertainty retainments are $r_{\textrm{ln}x}=0.54$ and $r_{\nu}=0.25$, $\tau=0.11$.}
\label{fig:in_0.5_out_0.5_cor_0_b_2_rewe}
\end{figure*}

The result for $b=2$ (high precision), $c = 0$ (uncorrelated data) and $\xi = 0.1$ is shown on figure \ref{fig:in_0.1_out_0.1_cor_0_b_2_rewe}, while  figure \ref{fig:in_0.5_out_0.5_cor_0_b_2_rewe} shows the result for $\xi = 0.5$ and the same other parameters. The right hand side plots show the  effect of reweighting in Ioffe time: we observe a large reduction of uncertainty, which remains effective far outside of the range of the data (the light orange zone) although the case $\xi = 0.5$ shows more fluctuations linked to the lesser coherence of the replica bundle. The average retainment of uncertainty in Ioffe time is $r_\nu = 0.16$ at $\xi = 0.1$, and $r_\nu = 0.25$ at $\xi = 0.5$.

On the other hand, the left hand side plots show a far less spectacular reduction of uncertainty in momentum space. The situation is actually inverted, with a larger reduction of uncertainty at $\xi = 0.5$ compared to $\xi = 0.1$. Indeed, the retainment of uncertainty remains very large at $r_{\textrm{ln}x} = 0.78$ at $\xi = 0.1$, whereas it is $r_{\textrm{ln}x} = 0.54$ at $\xi = 0.5$. The fact that the uncertainty is less reduced in momentum space compared to Ioffe time is an illustration of the imputation problem evoked in the introduction. The reconstruction of the $x$-dependence from limited Ioffe time data is an inverse problem which triggers a significant rise in uncertainty because the mock data produces no control over the ``high-frequency" behaviour of the momentum distribution. Therefore, reweighting is twice less efficient in momentum space compared to Ioffe time at $\xi = 0.5$, and five times less efficient at $\xi = 0.1$. Why does it behave so poorly at $\xi = 0.1$ in momentum space? The answer comes from the extreme coherence of the replicas of the GPD model in Ioffe time in the region where the reweighting is performed. This means that the GPD model contains in the end very little information in this region, such that the reweighting cannot efficiently select features of the distribution that would make a significant difference in momentum space.

We reiterate that the origin of the large coherence is the fact that positivity constrains tightly the GPD on a large part of the $x$ dependence at $\xi = 0.1$ (namely for $x > 0.1$). We add therefore lattice data in a region where another theoretical constraint in momentum space has already considerably reduced the freedom in modelling. Lattice data must be all the better to bear any impact that the initial modelling uncertainties are small. On the other hand, the model is much more flexible at $\xi = 0.5$, allowing lattice data a relatively better discriminating power in momentum space.

\begin{figure*}
\centering
\includegraphics[width=1\textwidth]{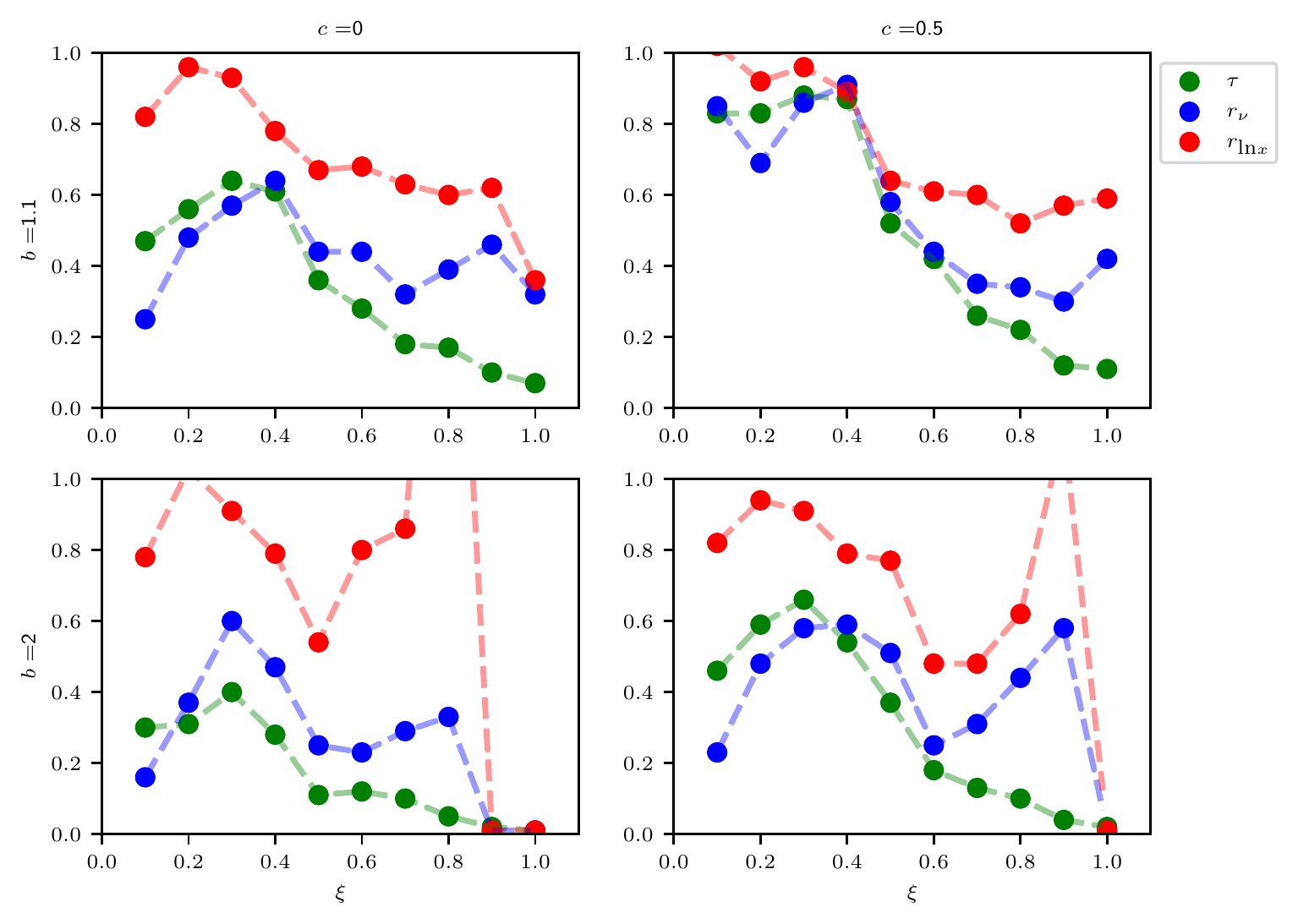}
\caption{Effective fraction of replicas retained after reweighting $\tau$ (green curve), retainment of uncertainty in Ioffe time (blue curve) and in momentum space (red curve) for the combinations of high and low noise (resp. $b = 1.1$ and $2$) and low and high correlation (resp. $c = 0$ and $0.5$).}
\label{fig:frac}
\end{figure*}

To compare the effect of reweighting at various values of $\xi$, we show in figure \ref{fig:frac} the effective fraction of surviving replicas $\tau$ and the retainment of uncertainty in Ioffe time and momentum space as functions of $\xi$.
As the value of $\xi$ increases, the replica bundle decoheres and the effective fraction of replicas $\tau$ drops quickly.
For $\xi > 0.7$, a full refit seems necessary due to the low statistics.
We observe that the reduction of uncertainty is systematically better in Ioffe time compared to momentum space, as expected due to the imputation problem. Using better and uncorrelated data ($b = 2, c= 0$) produces generally a significant reduction of uncertainty in Ioffe time compared to the other configurations, but that is not reflected in momentum space.
Overall, we observe in the case $b = 1.1$ (low precision) that uncertainty in momentum space decreases with larger values of $\xi$. The fact that uncertainty is so erratic for $b = 2$ has to do with the small value of $\tau$, which makes results unreliable at large $\xi$, but underlines how constraining the new data are compared to the prior model.

This highlights that with our modelling, it is in the large $\xi$ region, where positivity does not provide significant constraints on the GPD, that we would observe the largest effect due to the inclusion of lattice data. However, we have worked here at $t = 0$, where we could use the very well-known unpolarised PDF as an input to our model, and used a simplified version of the positivity constraint neglecting the role of the GPD $E$ for instance. In general, the behaviour of GPDs at $\xi = 0$ but $t \neq 0$ is one of the most crucial aspects of GPD phenomenology and can be accessed readily on the lattice. Although our current study invites to preferentially include data at large $\xi$, in a more complete setting, the extraction of more precise $t$-dependent PDFs plays a major role. We note that by studying both the limit $\xi = 0$ and large $\xi$, one handles two very interesting limits of GPDs: the first is directly involved in the hadron tomography program or the spin sum rule, whereas the latter probes a very different partonic dynamic in the hadron, with a sensitivity to pairs of partons. The small but non-zero values of $\xi$ can on the other hand be constrained rather efficiently by a combination of experimental data, positivity constraints and arguments from perturbation theory if working at a reasonably hard scale \cite{Dutrieux:2023qnz}. It remains however interesting to verify whether lattice data at moderately small value of $\xi$ are in good agreement with positivity constraints considering the words of caution on relying excessively on these inequalities at low renormalisation scale, as underlined for instance in \cite{Collins:2021vke}.

\begin{figure*}[!ht]
\centering
\includegraphics[width=1\textwidth]{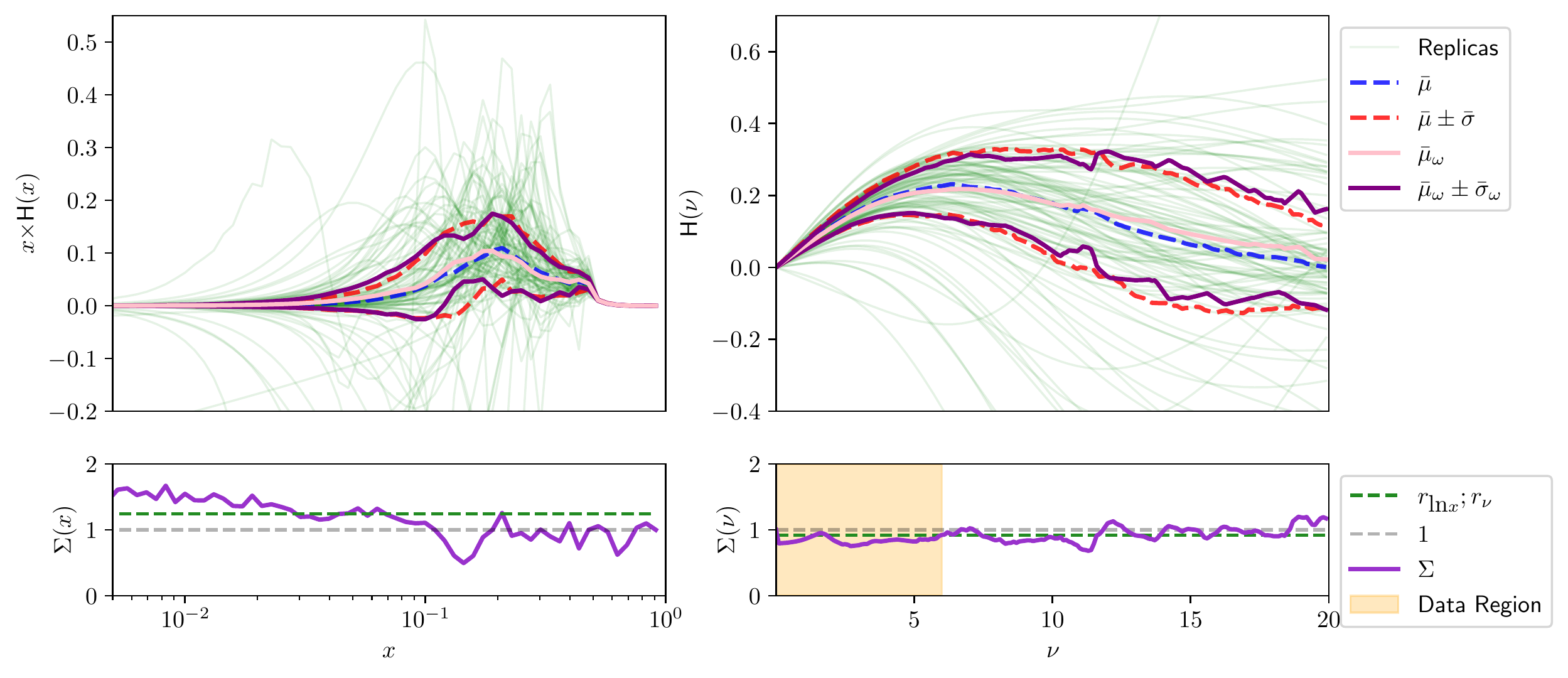}
\caption{Same caption as figure \ref{fig:in_0.1_out_0.1_cor_0_b_2_rewe} up to the fact that the GPD is shown at $\xi = 0.5$, with mock data added at $\xi = 0.1$, $b=1.1$ (low precision) and $c=0.5$ (correlated data). The average uncertainty retainments are $r_{\textrm{ln}x}=1.15$ and $r_{\nu}=0.93$, $\tau=0.83$.}
\label{fig:in_0.1_out_0.5_cor_0.5_b_1.1_rewe}
\end{figure*}

\subsection{Multikinematic Reweighting}

We would like to explore one last aspect, namely how reweighting is propagated to other values of $\xi$. So far, we have only studied the effect of reweighting exactly at the same value of $\xi$ as the data we were simulating. We show in figure \ref{fig:in_0.1_out_0.5_cor_0.5_b_1.1_rewe} the result of a reweighting where the data is added at $\xi = 0.1$, but we observe the impact on the GPD at $\xi = 0.5$, with $b=1.1$ (low precision) and $c = 0.5$ (correlated data). With these large uncertainties, the reweighting does not give any significant reduction of uncertainty at $\xi = 0.5$ in Ioffe time ($r_\nu = 0.93$), and even an increase of uncertainty in momentum space ($r_{\textrm{ln}x}=1.15$) by smearing the distribution. 

\begin{figure*}[!ht]
\centering
\includegraphics[width=1\textwidth]{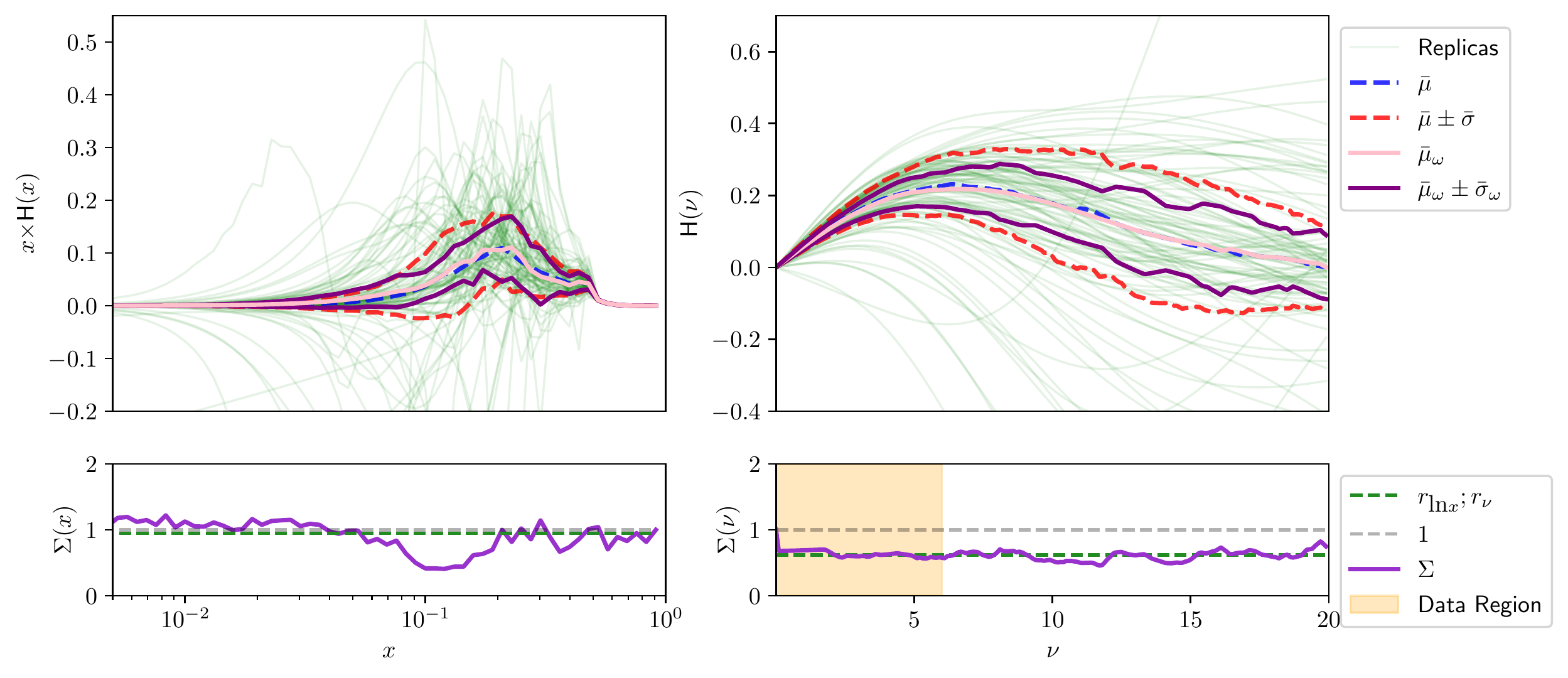}
\caption{Same caption as figure \ref{fig:in_0.1_out_0.1_cor_0_b_2_rewe} up to the fact that the GPD is shown at $\xi = 0.5$, with mock data added at $\xi = \{0.1, 0.2, 0.3\}$, $b=1.1$ (low precision) and $c=0.5$ (correlated data).
The average uncertainty retainments are $r_{\textrm{ln}x}=1.0$ and $r_{\nu}=0.82$, $\tau=0.77$.}
\label{fig:in_0.1_0.2_0.3_out_0_cor_0.5_b_1.1_rewe}
\end{figure*}

\begin{figure*}
\centering
\includegraphics[width=1\textwidth]{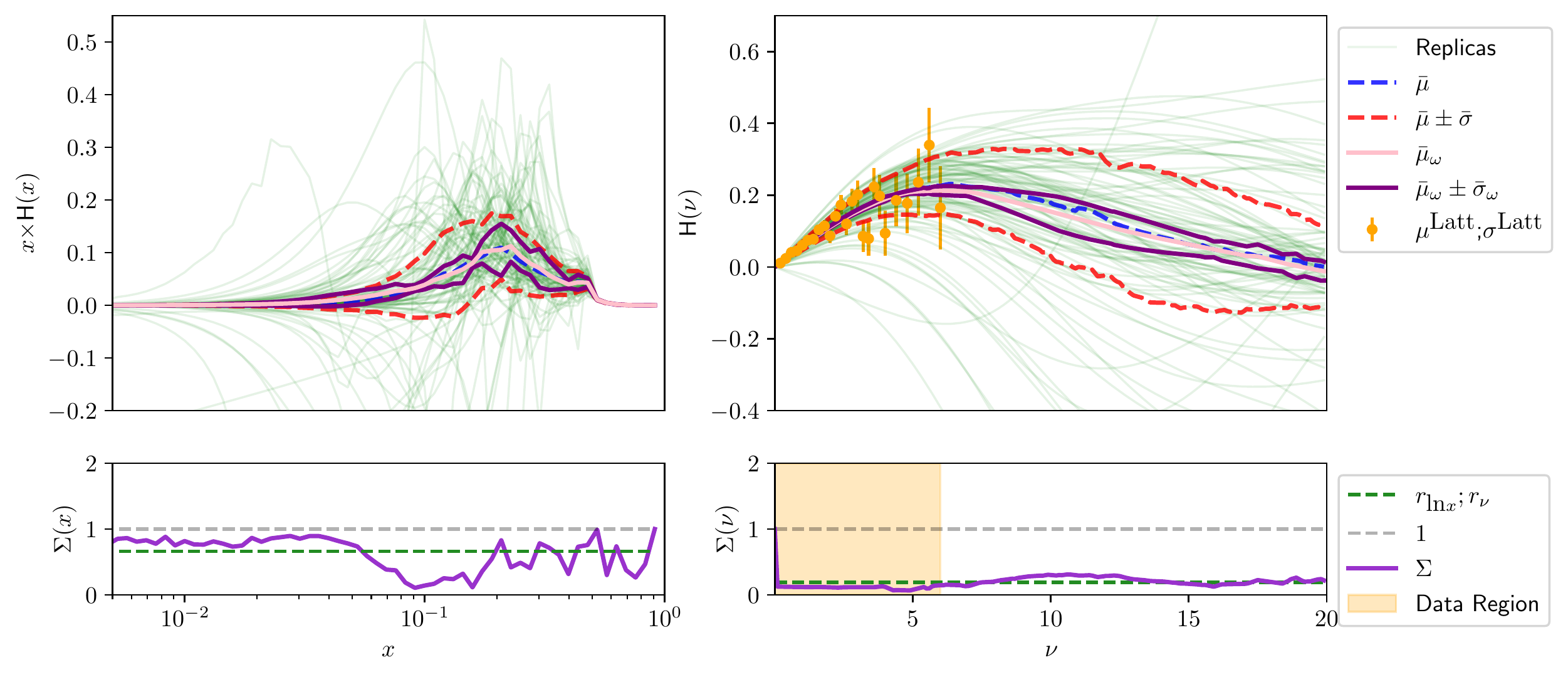}
\caption{Same caption as figures \ref{fig:in_0.1_out_0.1_cor_0_b_2_rewe} up to the fact that the GPD is shown at $\xi = 0.5$, with mock data added at $\xi = \{0.1, 0.2, 0.3, 0.4, 0.5\}$, $b=1.1$ (low precision) and $c=0.5$ (correlated data).
The average uncertainty retainments are $r_{\textrm{ln}x}=0.75$ and $r_{\nu}=0.65$, $\tau=0.57$.}
\label{fig:in_0.1_0.2_0.3_0.4_0_out_0.5_cor_0.5_b_1.1_rewe}
\end{figure*}

Let us now add data for $\xi \in \{0.1, 0.2,0.3\}$ while keeping $b=1.1$ and $c=0.5$.
The retainment of uncertainty at $\xi = 0.5$ drops to $r_\nu = 0.82$ and $r_x = 1.0$ (figure \ref{fig:in_0.1_0.2_0.3_out_0_cor_0.5_b_1.1_rewe}).
If we now add data for $\xi \in \{0.1, 0.2,0.3,0.4,0.5\}$, uncertainty retainment at $\xi = 0.5$ tightens to $r_\nu = 0.65$ and $r_x = 0.75$ (figure \ref{fig:in_0.1_0.2_0.3_0.4_0_out_0.5_cor_0.5_b_1.1_rewe}).
It is however not better than a simple reweighting directly at $\xi = 0.5$ which results in $r_\nu = 0.58$ and $r_x = 0.64$ (see Table \ref{tab:results} for all results, including various combinations of uncertainty and correlation that we have not mentioned in the text or figures).
This demonstrates that adding some data at a given value of $\xi$ only produces a rather minimal effect on other values of higher $\xi$ within our modelling of GPDs.

\begin{table*}[!ht]
\begin{centering}
\begin{tabular*}{\linewidth}{@{\extracolsep{\fill}} ccccccccc} \toprule
&Data& &&&&&Results& \\  
$\xi_{\text{Used}}$ & Precision & Correlation && $\xi_{\text{Shown}}$ && $\tau$ & $r_{\nu}$ & $r_{\textrm{ln}x}$ \\ \midrule
0.1 & Low & Low && 0.1 / 0.5 && 0.47 & 0.25 / 0.92 & 0.82 / 1.24 \\ 
0.1 & Low & High && 0.1 / 0.5 && 0.83 & 0.85 / 0.93 & 1.02 / 1.15 \\ 
0.1 & High & Low && 0.1 / 0.5 && 0.30 & 0.16 / 0.90 & 0.78 / 1.08 \\  
0.1 & High & High && 0.1 / 0.5 && 0.46 & 0.23 / 0.91 & 0.82 / 1.23 \\  
0.5 & Low & Low && 0.5 && 0.36 & 0.44 & 0.67 \\  
0.5 & Low & High && 0.5 && 0.52 & 0.58 & 0.64 \\ 
0.5 & High & Low && 0.5 && 0.11 & 0.25 & 0.54 \\ 
0.5 & High & High && 0.5 && 0.37 & 0.51 & 0.77 \\  
0.1 0.2 0.3 & Low & Low && 0.5 && 0.30 & 0.62 & 0.95 \\  
0.1 0.2 0.3 & Low & High && 0.5 && 0.77 & 0.82 & 1.00 \\
0.1 0.2 0.3 & High & Low && 0.5 && 0.10 & 0.34 & 0.54 \\ 
0.1 0.2 0.3 & High & High && 0.5 && 0.30 & 0.61 & 0.73 \\ 
0.1 0.2 0.3 0.4 0.5 & Low & Low && 0.5 && 0.16 & 0.19 & 0.66 \\  
0.1 0.2 0.3 0.4 0.5 & Low & High && 0.5 && 0.57 & 0.65 & 0.75 \\  
0.1 0.2 0.3 0.4 0.5 & High & Low && 0.5 && 0.03 & 0.13 & 0.45 \\ 
0.1 0.2 0.3 0.4 0.5 & High & High && 0.5 && 0.18 & 0.25 & 0.77 \\ \bottomrule 
\end{tabular*}
\end{centering}
\caption{Results as a function of the reweighting parameters.
Low Correlation: $c=0$, High Correlation: $c=0.5$, Low Precision: $b=1.1$, High Precision: $b=2$, $r_{\textrm{ln}x}$: Average uncertainty retainment in $x$, $r_{\nu}$: Average uncertainty retainment in $\nu$, $\tau$: Effective fraction of replicas retained after reweighting.}
\label{tab:results}
\end{table*}

%%% Local Variables:
%%% mode: latex
%%% TeX-master: "main"
%%% End:

%% file: sec_conclusion.tex
We have presented a study of the impact of mock lattice QCD data at moderate value of $\xi$ on a GPD model. The latter is based on machine learning techniques and fitted to the forward limit and diagonal $x = \xi$ of the phenomenological GK model, which represents the typical experimental information available on GPDs. We further enforce a positivity constraint, which considerably limits the freedom of the model in the region $x > \xi$. We observe as a result that our model uncertainties are largely autocorrelated in the small Ioffe-time region at small $\xi$, meaning that lattice data only bring minimal additional reduction of uncertainty in momentum space. We observe that the reduction of uncertainty in momentum space is systematically inferior to that in Ioffe time space, as a consequence of the inverse problem of relating the two representations of GPDs. We also observe that the addition of data at some low values of $\xi$ impacts only minimally the GPD at higher values of $\xi$.
However, the latter point happens in a context where the $t$-dependence is neglected. When restored, one recovers the $t$-dependent PDF at zero skewness, a quantity which is not directly constrained by experimental data. We thus expect that the impact of lattice QCD data in the low-$\xi$ region will increase with the value of $|t|$.

The Bayesian method employed here appears to be an adequate way to combine both experimental and lattice knowledge on GPDs, when the lattice data is globally in agreement with the prior model. However, our study highlighted the impact of correlations within the lattice data on a potential joint extraction. Real lattice data frequently present a very high degree of correlation, along with systematic effects that are only starting to be under control. This mandates a very careful treatment to avoid significant biases in the assessment of uncertainty reduction, which represents a challenge for the community.

%%% Local Variables:
%%% mode: latex
%%% TeX-master: "main"
%%% End: